\documentclass[a4paper,12pt,titlepage]{article}

\usepackage{amsmath,amssymb}
\usepackage{geometry}
\usepackage{hyperref}
\usepackage{braket}
\usepackage{titlesec}
\usepackage{hyperref}
\usepackage{fancyhdr}
\usepackage[dvipdfmx]{graphicx}
\usepackage{slashed}

\hypersetup{
    setpagesize=false,
    bookmarksnumbered=true,
    bookmarksopen=true,
    colorlinks=true,
    linkcolor=blue,
    citecolor=blue,
}

\begin{document}

\fancypagestyle{foot}
{
    \fancyfoot[L]{\(^{*}\)E-mail address: yakkuru$\_$111@ruri.waseda.jp}
    \fancyfoot[C]{}
    \fancyfoot[R]{}
    \renewcommand{\headrulewidth}{0pt}
    \renewcommand{\footrulewidth}{0.5pt}
}

\renewcommand{\footnoterule}{%
  \kern -3pt
  \hrule width \columnwidth
  \kern 2.6pt}


\begin{titlepage}
    \begin{flushright}
        \begin{minipage}{0.2\linewidth}
            \normalsize
            WU-HEP-19-01
        \end{minipage}
    \end{flushright}

    \begin{center}
        \vspace*{5truemm}
        \Large
        \bigskip \bigskip
        \LARGE \textbf{Matrix model and Yukawa couplings \\ on the noncommutative torus} 
        
        \Large

        \bigskip \bigskip
        Masaki Honda$^{1,*}$ 

        \smallskip

        {\large $^{1}$ {\it Department of Physics, Waseda University, Tokyo 169-8555, Japan}} \\ 

        \bigskip \bigskip \bigskip
        
        \large \textbf{Abstract}
    \end{center}
        The IKKT model is proposed as a non-perturbative formulation of superstring theory. We propose a Dirac operator on the noncommutative torus, which is consistent with the IKKT model, based on noncommutative geometry. Next, we consider zero-mode equations of the Dirac operator with magnetic fluxes. We find that zero-mode solutions have the chirality and the generation structures similar to the commutative case. Moreover, we compute Yukawa couplings of chiral matter fields.
    \thispagestyle{foot}
    
\end{titlepage}

\baselineskip 7.5mm


\tableofcontents
\clearpage

\section{Introduction}
Superstring theory is a promising candidate for a unified theory of all forces in nature. Superstring theory is only defined perturbatively and has infinite degenerate vacua. Therefore, it is said that superstring theory has no predictions for our world, and we need a more fundamental theory.

Matrix models are proposed as a non-perturbative formulation of superstring theory. In this paper, we focus on Ishibashi-Kawai-Kitazawa-Tsuchiya (IKKT) model $\cite{Ishibashi:1996xs}$. This model is derived from the matrix regularization of the Green-Schwarz action in the Schild gauge or large-N reduced model of ten-dimensional (10D) super Yang-Mills (SYM) theory. 

In the IKKT model, matter fields and degree of freedom of spacetime are considered to be embedded in matrices. Several attempts have been made to show it. In Ref. $\cite{Chatzistavrakidis:2011gs}$, the authors considered an intersecting flat D-branes system based on the IKKT model. To analyze the chirality and the generation structures of the system, they used the analogy of the harmonic oscillator in quantum mechanics. They showed the existence of a chiral zero-mode which are coming from a string connecting two different D-branes. However, they considered a simple configuration, e.g., two D-branes are orthogonal. Therefore, they could not realize multiple chiral zero-modes. In addition, in more general configurations, we cannot easily find the chiral zero-modes by using the same method because the complicated contribution comes from the mixing term between two D-branes. In Ref. $\cite{Steinacker:2014fja}$, the authors considered intersecting fuzzy spheres as compact D-branes and realized two chiral zero-modes. To admit a fuzzy sphere as a classical solution, we must introduce a new term. However, the origin of such a term is not clear. In either case, it is difficult to obtain the number of chiral zero-modes that we expect. In Ref. $\cite{Aoki:2014cya}$, the authors challenged the realization of three generations by using numerical analysis. By considering the squashed fuzzy sphere in addition to the fuzzy spheres, they succeeded in realizing three generations numerically. On the other hand, the authors of Ref. $\cite{Aoki:2008ik}$, considered a fuzzy torus with a magnetic flux based on the finite-matrix formulation of gauge theories\footnote{In this paper, we use {\it noncommutative} to describe infinite dimensional representations or operator algebras and {\it fuzzy} to describe finite dimensional representations or approximations.}. They computed the index of the overlap Dirac operator on the fuzzy torus by the Monte Carlo simulations and showed the numerical results that are consistent with the index theorem. In this formulation, the exact relationship with the IKKT model is not clear because we must introduce the special type of the Dirac operator to obtain the non-trivial index of the Dirac operator. Moreover, it is still difficult to compute physically important quantities like Yukawa couplings.

The purposes of this paper are to define a Dirac operator that is consistent with the IKKT model, to analyze the chirality and the generation structures and to compute Yukawa couplings of chiral matter fields.

To analyze concretely, we consider the noncommutative (NC) torus as a classical solution of the IKKT model. It is known that the NC torus as an irrational rotation ring can be realized in the IKKT model $\cite{Connes:1997cr}$. In addition, we consider the analogy of the toroidal compactifications of 10D SYM theory. In Ref. $\cite{Cremades:2004wa}$, the authors considered the toroidal compactifications of 10D SYM theory with magnetic fluxes. A key concept is the twisted bundle. The twisted bundle can be interpreted as a compatibility condition between the periodic boundary conditions on the torus and the gauge transformations. This compatibility condition implies that the magnetic fluxes are quantized, then the zero-mode solution of the Dirac equation can have the chirality and the degeneracies. The authors of Ref. $\cite{Cremades:2004wa}$ identified these results as the chirality and the generation structures in the four-dimensional effective theory and computed the Yukawa couplings by computing the overlap integral over the torus. From the viewpoint of phenomenology in superstring theory, this result is very important. Therefore, we consider the twisted bundle on the NC torus.

In the following, we propose a Dirac operator on the NC torus that is consistent with the IKKT model, and we solve the zero-mode equation of this Dirac operator with magnetic fluxes. Then, we compute the normalization factors of the zero-mode solutions and Yukawa couplings of chiral matter fields. 

The organization of this paper is as follows. In section 2, we briefly review the IKKT model and the realization of the NC torus in the IKKT model. In section 3, we review the basic results of the toroidal compactifications of 10D SYM theory. In section 4, we introduce differential operators on the NC torus based on noncommutative geometry. In addition, we propose a Dirac operator that is consistent with the IKKT model. In section 5, we solve the zero-mode equation by using the analogy of the Fourier transformation. In addition, we compute the normalization factors of the zero-mode solutions and the Yukawa couplings of chiral matter fields. To compute, we define the trace that is consistent with the gauge transformations and the torus translations. Section 6 contains conclusions and discussion. 

\section{IKKT model}
\subsection{Ishibashi-Kawai-Kitazawa-Tsuchiya (IKKT) model}
The action of the IKKT model is defined as follows $\cite{Ishibashi:1996xs}$,

\begin{align}
  \label{action of IKKT model}
  S=-\frac{1}{g^{2}} \text{Tr} \left( \frac{1}{4} [X_{M},X_{N}] [X^{M}, X^{N}] +\frac{1}{2} \bar{\psi} \Gamma^{M} [X_{M},\psi] \right),
\end{align}
where $X^{M}$ $(M=0 \sim 9)$ is a 10D vector and $N \times N$ Hermitian matrix, $\psi$ is a 10D Majorana-Weyl spinor whose components are $N \times N$ matrices and $g$ is a scale factor. Indices are contracted by the Minkowski metric. On the other hand, we will see that this action admits an infinite dimensional representation like linear operators in the next subsection.

The action (\ref{action of IKKT model}) has some symmetries:

\begin{align}
  &\delta^{(1)}X^{M} = i \bar{\epsilon}_{1} \Gamma^{M} \psi, \quad \delta^{(1)} \psi = \frac{i}{2} \Gamma^{M N} [X_{M},X_{N}] \epsilon_{1}, \label{susy transformation 1} \\
  &\delta^{(2)}X^{M} = 0, \quad \delta^{(2)}\psi = \epsilon_{2} 1_{N}, \label{susy transformation 2} \\
  &\delta_{T}X^{M}=c^{M}1_{N}, \quad \delta_{T}\psi=0, \label{shift symmetry} \\
  &\delta_{G}X^{M}=i[\Lambda,X^{M}], \quad \delta_{G}\psi=i[\Lambda,\psi],\label{gauge symmetry}
\end{align}
where $\epsilon_{i}$ $(i=1,2)$ is 10D Majorana-Weyl spinor as a Grassmann odd parameter, $c^{M}$ is a 10D constant vector, $1_{N}$ is the $N \times N$ identity matrix, and $\Lambda$ is a $N \times N$ Hermitian matrix. 

Eqs. (\ref{susy transformation 1}) and (\ref{susy transformation 2}) can be identified $\mathcal{N}=2$ supersymmetry (SUSY) as follows. If we consider a linear combination of $\delta^{(1)}$ and $\delta^{(2)}$ as

\begin{align*}
  \tilde{ \delta }^{(1)} \equiv \delta^{(1)}+\delta^{(2)}, \quad \tilde{\delta}^{(2)} \equiv i(\delta^{(1)} - \delta^{(2)}),
\end{align*}
then we can obtain

\begin{align}
  [ \tilde{\delta}^{(i)}_{\epsilon}, \tilde{\delta}^{(j)}_{\xi}]X^{M}=2i \bar{\epsilon} \Gamma^{M} \xi \delta^{ij}, \quad [ \tilde{\delta}^{(i)}_{\epsilon}, \tilde{\delta}^{(j)}_{\xi}]\psi=0 \quad (i,j=1,2), \label{susy true}
\end{align}
where we used eq. (\ref{gauge symmetry}) and the equation of motion of $\psi$. If we identify $X^{M}$ as a 10D spacetime coordinate, eq. (\ref{susy true}) is $\mathcal{N}=2$ on-shell SUSY algebra. In this sense, we often say that the degree of freedom of spacetime is embedded in matrices. From this, there is a possibility of analysis for the dynamics of spacetime, e.g., a mechanism of compactifications. 

When $\psi=0$, the equation of motion of $X^{M}$ is 

\begin{align}
  \label{eom of X}
  [X^{N}, [X^{M}, X^{N}] ]=0.
\end{align}
The simplest solution is $^{\forall}[X^{M},X^{N}]=0$. However, we can show that attractive forces act between the eigenvalues in the one-loop effective potential around this vacuum. Therefore, they do not spread, and the correspondence with the original theory does not hold. In this case, we need more conditions on gauge groups \cite{Austing:2001bd,Austing:2001pk}.

The second simplest, however, an interesting solution is that some commutators are proportional to the identity matrix, i.e.,

\begin{align}
  \label{second simplest solution}
  [X^{M},X^{N}]=i\theta^{M N},   
\end{align}
where $\theta^{M N}$ is a real anti-symmetric matrix with respect to the Lorentz indices. We omit the identity matrix. When each matrix has a finite size, eq. (\ref{second simplest solution}) is a contradiction. We should interpret eq. (\ref{second simplest solution}) as it is satisfied at large matrix size. This is the correspondence between a function algebra and a matrix algebra in the matrix regularization. On the other hand, the infinite dimensional representation, which we are interested, strictly respects eq. (\ref{second simplest solution}).

We can also interpret eq. (\ref{second simplest solution}) as a D-branes (BPS) configuration as follows. We can find that the transformation by $\delta^{(1)}$ is proportional to the identity matrix and cancel the transformation by $\delta^{(2)}$. Explicitly, if we set $\epsilon_{2}=\pm \frac{1}{2} \theta^{M N} \Gamma_{M N} \epsilon_{1}$, then 

\begin{align*}
  (\delta^{(1)} \pm \delta^{(2)})X^{M}=0, \quad  (\delta^{(1)} \pm \delta^{(2)}) \psi=0.
\end{align*}
Therefore, the half of $\mathcal{N}=2$ SUSY is preserved in this background, and the one-loop effective potential exactly vanishes due to SUSY. 
It is an advantage of this model that a many-body system of D-branes can be realized. When we consider two sets of eq. (\ref{second simplest solution}),

\begin{align*}
  [X^{M}_{(1)}, X^{N}_{(1)}]=i\theta^{M N}_{(1)}, \quad [X^{M}_{(2)}, X^{N}_{(2)}]=i\theta^{M N}_{(2)},
\end{align*}
then 

\begin{align}
  \label{multi D-brane solution}
  X^{M}=
  \begin{pmatrix}
    X^{M}_{(1)} & 0\\
    0 & X^{M}_{(2)}
  \end{pmatrix}
\end{align}
is also a solution of eq. (\ref{eom of X}). In general, eq. (\ref{multi D-brane solution}) is not a D-branes configuration. However, some configurations become a D-branes configuration again, e.g., the case of $X^{M}_{(1)}=X^{M}_{(2)}$ is interpreted as two coincident D-branes, and other cases are considered in Ref. \cite{Ishibashi:1996xs}.

\subsection{NC torus in the IKKT model}
In this subsection, we briefly review the NC torus in the IKKT model based on Refs. \cite{Connes:1997cr,Konechny:2000dp}. A unitary transformation is defined by

\begin{align}
  \label{unitary transformation}
  X^{M} \rightarrow U X^{M} U^{-1}, \quad \psi \rightarrow U \psi U^{-1},
\end{align}
where $U$ is a unitary matrix (operator). If the trace in the action (\ref{action of IKKT model}) has the cyclic property, this unitary transformation becomes a symmetry of the IKKT model. The infinitesimal form of eq. (\ref{unitary transformation}) corresponds to eq. (\ref{gauge symmetry}). The authors of Ref. \cite{Connes:1997cr} showed that the NC torus can be realized by this unitary transformation.

We restrict the action (\ref{action of IKKT model}) to the subspace of $(X^{M},\psi)$ belonging to the same gauge class before and after translations in the directions $X^{4}$ and $X^{5}$. In other word, we consider conditions as follows, 

\begin{align}
  \label{compactification conditions}
  &U_{4} X^{4} U^{-1}_{4}=X^{4}+2\pi R_{4}, \quad U_{5} X^{5} U^{-1}_{5} =X^{5}+2\pi R_{5}, \notag \\
  &U_{i} X^{M} U^{-1}_{i} =X^{M} \quad (i \neq M, i=4,5,M=0 \sim 9 ), \notag\\
  &U_{i} \psi U^{-1}_{i}=\psi \quad (i=4,5),
\end{align}
where $R_{i}$ $(i=4,5)$ is the scalar matrix (operator) of a real coefficient corresponding to the periods of the torus. In the following, let us define $M=0 \sim 9$, $\mu=0 \sim 3$ and $i=4 \sim 9$.

We can see that any finite dimensional representations do not satisfy the conditions (\ref{compactification conditions}). However, we can find solutions if we interpret $X^{M}$ and $\psi$ as operators on an infinite dimensional Hilbert space\footnote{In general, the trace on an infinite dimensional Hilbert space does not satisfy the cyclic property (e.g., the position operators, the momentum operators and their commutation relations in quantum mechanics). We will define the trace which is consistent with the gauge transformations (\ref{unitary transformation}) and the conditions (\ref{compactification conditions}).}. 

From eq. (\ref{compactification conditions}), we can confirm that $U_{4}U_{5}U^{-1}_{4}U^{-1}_{5}$ commutes with $X^{M}$ and $\psi$. Therefore, we assume (or we may be able to apply Schur's lemma for countable dimensional cases \cite{Dixmier1963,Quillen1969} because we can also realize $U_{4}$ and $U_{4}$ as operators on $l^{2}(\mathbb{R})$) that $U_{4}U_{5}U^{-1}_{4}U^{-1}_{5}$ is a scalar operator, i.e., 

\begin{align}
  \label{NC torus algebra}
  U_{4}U_{5} =e^{2 \pi i \theta} U_{5} U_{4},
\end{align}
where $\theta$ is a real parameter. Eq. (\ref{NC torus algebra}) is the algebra of the NC torus in mathematics. Precisely, we can restrict $\theta \in [0,1/2]$ because of some isomorphisms. Depending on whether the parameter $\theta$ is a rational number or an irrational number, the mathematical structure differs. In fact, the algebra of the NC torus admits a finite dimensional representation if $\theta \in \mathbb{Q}$, and we cannot realize such algebra by the finite dimensional representation if $\theta \in \mathbb{R} \backslash \mathbb{Q}$. In this paper, we assume $\theta \in \mathbb{R} \backslash \mathbb{Q}$ because we are interested in the infinite dimensional representation of the NC torus. In this sense, $X^{M},\psi,U_{4}$ and $U_{5}$ are no longer finite dimensional matrices but operators. We use ``hat'' to indicate that it is an operator (we omit the identity operator).

In the following, we show a concrete representation of $\hat{X}^{M},\hat{U}_{4}$ and $\hat{U}_{5}$. Let us start from a Hilbert space $\mathcal{H}=L^{2}(\mathbb{R}) \otimes \mathbb{C}^{m}$, $m \in \mathbb{N}$ ($m$ corresponds to $\mathbb{Z}_{m}$ in the usual construction \cite{Connes:1997cr,Konechny:2000dp,Rieffel1988}). We can realize $\hat{U}_{4}$ and $\hat{U}_{5}$ as operators on $\mathcal{H}$, i.e., 

\begin{align}
  \label{action of U}
  \hat{U}_{4} \Ket{x:j} = \Ket{x-\frac{n}{m}+\theta : j-1}, \quad \hat{U}_{5} \Ket{x:j}=\exp \left[ -2 \pi i \left( x-\frac{nj}{m} \right) \right] \Ket{x:j},
\end{align}
where $n$ is some integer. For simplicity, we assume that $m$ and $n$ are positive and co-prime each other (or $m=1$ and $n=0$). We define $\hat{X}^{4}$ and $\hat{X}^{5}$ as 

\begin{align}
  \label{action of X}
  \hat{X}^{4}:= \frac{2 \pi m}{n-m\theta}R_{4}\hat{x} \otimes 1_{m}, \quad \hat{X}^{5} := R_{5} \hat{p} \otimes 1_{m},
\end{align}
where $1_{m}$ is the $m \times m$ identity matrix, $\hat{x}$ and $\hat{p}$ are the position operator and the momentum operator on one-dimensional quantum mechanics (where $\hbar$=1), respectively. We can see that $\theta^{45}=\frac{2\pi m}{n-m \theta}R_{4}R_{5}$ and easily confirm that eqs. (\ref{compactification conditions}) and (\ref{NC torus algebra}) are satisfied. In the following, since the part acting on $\mathbb{C}^{m}$ is the identity matrix, we use the Hilbert space spanned by $\{ \Ket{X^{4}} ; \hat{X^{4}} \Ket{X^{4}}=X^{4}\Ket{X^{4}} \}$ or $\{ \Ket{X^{5}} ; \hat{X^{5}} \Ket{X^{5}}=X^{5}\Ket{X^{5}} \}$.

From an analogy of quantum mechanics,

\begin{align}
  \label{inner product}
  \Braket{X^{4}|X^{5}}=\frac{1}{A} \exp \left[ i \frac{X^{4}X^{5}}{\theta^{45}} \right], \quad A=(2\pi)^{2}R_{4}R_{5} \quad (\text{the area of the torus}),
\end{align}
since the periodicity is realized by the unitary operators $\hat{U}_{4}$ and $\hat{U}_{5}$, and we restrict the fundamental (or physical) region to $0 \leq X^{4} \leq 2\pi R_{4}$ and $0 \leq X^{5} \leq 2\pi R_{5}$\footnote{Let $\mathcal{H}_{m}$ $(m \in \mathbb{Z})$ be a Hilbert space such that the spectrum of the Hermitian operator $\hat{X}^{4}$ lies in the interval $2\pi m R_{4} \leq X^{4} \leq 2\pi (m+1)R_{4}$. In this case, $\mathcal{H}_{m}$ is unitary equivalent to $^{\forall}\mathcal{H}_{n \in \mathbb{Z}}$ Therefore, we interpret the fundamental (or physical) region with respect to the $X^{4}$-direction is $0 \leq X^{4} \leq 2\pi R_{4}$. We interpret the $X^{5}$-direction as well.}.

\section{Toroidal compactifications of 10D SYM theory}
In this section, we review the basic results of toroidal compactifications of 10D SYM theory based on Refs. \cite{Cremades:2004wa,Tenjinbayashi:2005sy}.

\subsection{Twisted bundle on the torus}
\label{3.1}
Let us start from $U(1)$ gauge theory on the torus $T^{2}$. In this paper, we take a lattice $\Lambda \simeq \mathbb{Z}^{2}$ that is generated by $v^{1}=(2\pi R_{4},0)$ and $v^{2}=(0,2\pi R_{5})$, and $T^{2}$ is defined by $\mathbb{R}^{2} \slash \Lambda$.

A constant magnetic flux is introduced by 

\begin{align}
  \label{gauge background on torus}
  A_{4}=0, \quad A_{5}=Fx^{4},
\end{align}
then the magnetic flux $F_{45}=F$. We use the same gauge choice ({\it axial gauge} in Ref. \cite{Tenjinbayashi:2005sy}) in section \ref{5}.

We must confirm that the gauge theory is well-defined on $T^{2}$. Obviously, we should confirm only about $x^{4}$-direction. The covariant derivatives transform like

\begin{align}
  \label{translation of covariant derivatives}
  D_{4} \rightarrow D_{4}, \quad D_{5} \rightarrow D_{5}-2\pi i R_{4} F.
\end{align}
If this translation can be absorbed as the gauge symmetry, $U(1)$ gauge theory on $T^{2}$ is well-defined. Actually, we can realize eq. (\ref{translation of covariant derivatives}) by 

\begin{align*}
  \Omega_{4}(x^{4},x^{5}) D_{4} \Omega^{-1}_{4}(x^{4},x^{5})=D_{4}, \quad \Omega_{4}(x^{4},x^{4}) D_{5} \Omega^{-1}_{4}(x^{4},x^{5})=D_{5}-2\pi i R_{4} F,
\end{align*}
where $\Omega_{4}(x^{4},x^{5})=\exp [2\pi i R_{4} F x^{5}]$. In general, we need to consider another gauge transformation $\Omega_{5}(x^{4},x^{5})$ associated with the translation along the $x^{5}$-direction. In the above case, $\Omega_{5}(x^{4},x^{5})=1$. The above concept is called the twisted bundle.

In addition, we must confirm a consistency condition

\begin{align}
  \label{consistency condition on torus}
  \Omega_{5}(x^{4}+2\pi R_{4},x^{5}) \Omega_{4}(x^{4},x^{5})=\Omega_{4}(x^{4},x^{5}+2\pi R_{5}) \Omega_{5}(x^{4},x^{5}).
\end{align}
Eq. (\ref{consistency condition on torus}) implies 

\begin{align}
  \label{quantization of magnetic flux}
  \frac{A}{2\pi }F \in \mathbb{Z},
\end{align}
where $A=(2\pi)^{2}R_{4}R_{5}$ is the area of the torus. Namely, the magnetic flux on the torus is quantized.

\subsection{Zero-modes of Dirac operator on $T^{2}$}
We can analytically solve the zero-mode equations of the Dirac operator with the background gauge field (\ref{gauge background on torus}). The Dirac operator is defined by 

\begin{align*}
  \slashed{D}=\sum_{i=4,5} \sigma_{i} D_{i}=
  \begin{pmatrix}
    0 & \partial_{4}-i\partial_{5}-Fx^{4} \\
    \partial_{4}+i\partial_{5}+Fx^{4} & 0
  \end{pmatrix},
\end{align*}
where $\sigma_{4}$ and $\sigma_{5}$ are the Pauli matrices $\sigma_{1}$ and $\sigma_{2}$, respectively. For convenience, we rewrite the constant magnetic flux as $F=\frac{2\pi }{A} N \nu$, where $N$ is a positive integer and $\nu=\text{sign}(F)$. The zero-mode equation of the Dirac operator is

\begin{align*}
  \slashed{D}\psi=0.
\end{align*}
If $\psi$ is labeled by the eigenvalue of the chirality matrix $\sigma_{3}$, i.e.,

\begin{align}
  \label{chirality assignment}
  \psi=
  \begin{pmatrix}
    \psi^{+} \\
    \psi^{-}
  \end{pmatrix},
\end{align}
the zero-mode equations of each component can be written as 

\begin{align}
  \label{zero-mode equation each component on torus}
  (\partial_{4}+is\partial_{5}+sFx^{4})\psi^{s}=0 \quad (s=\pm 1).
\end{align}
We can easily factorize $\psi^{s}$ satisfying eq. (\ref{zero-mode equation each component on torus}) as

\begin{align*}
  \psi^{s}(x^{4},x^{5})=\exp \left[ -\frac{1}{2}sF (x^{4})^{2} \right] f^{s}(x^{4}+isx^{5}),
\end{align*}
where $f^{s}(x^{4}+isx^{5})$ satisfies

\begin{align}
  \label{pbc function f}
  &f^{s}(x^{4}+2\pi R_{4}+isx^{5})= \exp \left[ \frac{N}{R_{5}} s\nu (x^{4}+isx^{5}+\pi R_{4}) \right] f^{s}(x^{4}+isx^{5}),\notag \\
  &f^{s} \left(x^{4}+is(x^{5}+2\pi R_{5})  \right) =f^{s}(x^{4}+isx^{5}).
\end{align}
$f^{s}(x^{4}+isx^{5})$ is periodic with respect to the $x^{5}$-direction. Therefore, we can use the Fourier expansion as
\begin{align*}
  f^{s}(x^{4}+isx^{5})=\sum_{n \in \mathbb{Z}} C^{s}_{n} \exp \left[\frac{ns}{R_{5}}(x^{4}+isx^{5}) \right].
\end{align*}
From the first condition of eq. (\ref{pbc function f}), the Fourier coefficients satisfy

\begin{align}
  \label{Fourier coefficient condition} 
  C^{s}_{n}=\exp \left[ \frac{s \pi R_{4}}{R_{5}} (N\nu -2n) \right] C^{s}_{n-N\nu}.
\end{align}
If we decompose $n$ into $n=Np+q$ $(p \in \mathbb{Z}, q=0,...,N-1)$, thus we can rewrite eq. (\ref{Fourier coefficient condition}) as

\begin{align*}
  C^{s}_{Np+q}=\exp \left[-\frac{\pi R_{4}}{R_{5}} s\nu (Np^{2}+2pq)  \right]C^{s}_{q}.
\end{align*}
From the above, $f^{s}(x^{4}+isx^{5})$ can be written as

\begin{align*}
  f^{s}(x^{4}+isx^{5})= \sum_{q=0}^{N-1} C^{s}_{q} \sum_{p \in \mathbb{Z}} \exp \left[-\frac{\pi R_{4}}{R_{5}} s\nu (Np^{2}+2pq) \right] \exp \left[\frac{s}{R_{5}}(Np+q)(x^{4}+isx^{5}) \right].
\end{align*}
If $\nu s =+1$, $f^{s}(x^{4}+isx^{5})$ converges. In other words, the sign of the magnetic flux determines the chirality of the zero-mode solution. In addition, the summation with respect to $q$ means that there are $N$-independent zero-mode solutions. In Ref. \cite{Cremades:2004wa}, these results are interpreted as the chirality and the generation structures in the four-dimensional effective theory.

Therefore, each degenerated solution of eq. (\ref{zero-mode equation each component on torus}) is 

\begin{align*}
  \psi^{s}_{q}(x^{4},x^{5})=C^{s}_{q} \exp \left[-\frac{|F|}{2} (x^{4})^{2} \right] \sum_{p \in \mathbb{Z}}  \exp \left[-\frac{\pi R_{4}}{R_{5}} s\nu (Np^{2}+2pq) \right] \exp \left[\frac{s}{R_{5}}(Np+q)(x^{4}+isx^{5}) \right].
\end{align*}
Each $C^{s}_{q}$ is determined by the normalization condition, i.e., 

\begin{align}
  \label{normalization condition torus}
  \int_{0}^{2\pi R_{4}} dx^{4} \int_{0}^{2\pi R_{5}} dx^{5} \psi^{s,*}_{\bar{q}} \psi^{s}_{q}=\delta_{\bar{q}q}.
\end{align}
From eq. (\ref{normalization condition torus}),

\begin{align}
  |C^{s}_{q}|^{2} \cdot 2\pi R_{5} \exp \left[\frac{2\pi R_{4}}{NR_{5}}q^{2}  \right] \sqrt{\frac{\pi}{|F|}}=1.
\end{align}
Eventually,

\begin{align}
  \label{normalized zero-mode solution torus}
  \psi^{s}_{q}(x^{4},x^{5})&=\left(2\pi R_{5} \sqrt{\frac{\pi}{|F|}} \right)^{-1/2} \exp \left[-\frac{|F|}{2} (x^{4})^{2} \right] \notag \\
  & \hspace{4cm} \times \sum_{p \in \mathbb{Z}}  \exp \left[-\frac{\pi R_{4}}{R_{5}} s\nu (Np^{2}+2pq) \right] \exp \left[\frac{s}{R_{5}}(Np+q)(x^{4}+isx^{5}) \right] \notag \\
  &=\left(2\pi R_{5} \sqrt{\frac{\pi}{|F|}} \right)^{-1/2} \exp \left[-\frac{|F|}{2} (x^{4})^{2} \right] \vartheta 
  \begin{bmatrix}
    q/N \\ 0
  \end{bmatrix}
  \left( \frac{sN}{2\pi i R_{5}}(x^{4}+isx^{5}),i\frac{R_{4}}{R_{5}}N \right),
\end{align}
where the function $\vartheta$ is the Jacobi theta-function

\begin{align*}
  \vartheta
  \begin{bmatrix}
    a \\ b
  \end{bmatrix}
  (\nu, \tau)
  =\sum_{p \in \mathbb{Z} } \exp \left[\pi i (a+p)^{2}\tau + 2\pi i (a+p)(\nu+b) \right],
\end{align*}
where $\text{Im}\tau>0$.

We can easily generalize the above discussion to non-Abelian, bi-fundamental or higher dimensional torus.

\section{Dirac operator on the NC torus and the IKKT model}
In this section, we introduce differential operators on the NC torus and construct them from the IKKT model.

\subsection{Differential operators on the NC torus}
Let $\mathcal{A}$ be a $C^{*}$-algebra over $\mathbb{C}$ and $d: \mathcal{A} \rightarrow \mathcal{A}$ be a linear map. $d$ is called a derivation on $\mathcal{A}$ if $d$ satisfies the Leibniz rule, i.e.,

\begin{align*}
  d(ab)=(da)b+a(db), \quad d(\lambda a)=\lambda (da),
\end{align*}
for all $a,b \in \mathcal{A},\lambda \in \mathbb{C}$.

In mathematics, the NC torus is the $C^{*}$-algebra generated by $\hat{U}_{4}$ and $\hat{U}_{2}$ satisfying eq. (\ref{NC torus algebra}). Therefore, derivations are completely defined by how they act on the two generators. We define

\begin{align}
  \label{basic derivations}
  \delta^{4} \hat{U}_{4}=\frac{i}{R_{4}} \hat{U}_{4}, \quad \delta^{5} \hat{U}_{5}=\frac{i}{R_{5}} \hat{U}_{5}, \quad \delta^{4}\hat{U}_{5}=\delta^{5}\hat{U}_{4}=0
\end{align}
and satisfy the linearity and the Leibniz rule. $\delta^{i}$ $(i=4,5)$ is called the basic derivation. The factors $\frac{1}{R_{4}}$ and $\frac{1}{R_{5}}$ come from the definition of the torus, c.f., subsection \ref{3.1}. We can confirm that the basic derivations commute each other.

The Dirac operator on the NC torus is defined by the basic derivations

\begin{align*}
  \slashed{D}:= i \sum_{i=4,5} \sigma_{i} \delta^{i},
\end{align*}
where $\sigma_{4}$ and $\sigma_{5}$ are the Pauli matrices $\sigma_{1}$ and $\sigma_{2}$, respectively. 

\subsection{Dirac operator on the NC torus based on the IKKT model}
\label{4.2}
In this subsection, we construct the differential operators, which are introduced in the previous subsection, based on the IKKT model. First, we focus on noncommutative SYM theory based on the IKKT model \cite{Steinacker:2007dq}.

We consider the expansion of the action (\ref{action of IKKT model}) around the specific background: spacetime $\hat{X}^{M}_{\text{bg}}$ satisfying eq. (\ref{second simplest solution}), gauge field $\hat{A}_{\text{bg}, M}$ will be defined in section \ref{5}. Namely,

\begin{align}
  \label{fluctuation}
  \hat{X}^{M}=\hat{X}^{M}_{\text{bg}}+\theta^{M N}(\hat{A}_{\text{bg},N}+\hat{A}_{N}).
\end{align}
We refer to Appendix \ref{A} for the details. A point is that the partial derivatives are defined as

\begin{align}
  \label{partial derivative NC torus}
  \partial_{M}:= -i(\theta^{-1})_{M N} \left[\hat{X}^{N}_{\text{bg}}, \cdot \right],
\end{align}
where $(\theta^{-1})_{M N}$ is the inverse matrix of $\theta^{M N}$. We can confirm that the partial derivatives satisfy $\partial_{M} \hat{X}^{N}_{\text{bg}}=\delta^{N}_{M}$. Here, we assume that $\theta^{M N}$ is non-degenerate. In the following, we identify $\hat{X}^{M}_{\text{bg}}$ with eq. (\ref{action of X}) (i.e., we mainly focus on the NC torus ($M=0 \sim 9 \rightarrow i=4,5$)), and we omit the subscription ``bg''.
 
The action of $\partial_{i}$ on the algebra of the NC torus is

\begin{align*}
  \partial_{i} \hat{U}_{j}&=-i (\theta^{-1})_{ik} \left[\hat{X}^{k}, \hat{U}_{j}\right]\\
  &=(\theta^{-1})_{ij} \cdot 2\pi i R_{j} \hat{U}_{j}.
\end{align*}
Therefore, the basic derivations on the NC torus based on the IKKT model are defined as

\begin{align}
  \label{basic derivations based on IKKT model}
  \delta^{i}:= \frac{1}{2\pi (R_{i})^{2}} \theta^{ij} \partial_{j}.
\end{align}
We can confirm that the basic derivations (\ref{basic derivations based on IKKT model}) satisfy eq. (\ref{basic derivations}). In addition, eq. (\ref{second simplest solution}) and the Jacobi identity assures the commutativity of the basic derivations (\ref{basic derivations based on IKKT model}).

From the above, we propose a Dirac operator on the NC torus that is consistent with the IKKT model as follows\footnote{The subscription ``phys'' means that the Dirac operator (\ref{Dirac operator consistent with IKKT model}) has the mass dimension 1. This mass dimension is the same with the usual Dirac operator in QFT.},

\begin{align}
  \label{Dirac operator consistent with IKKT model}
  \slashed{D}_{\text{phys}}:= 2\pi i \sum_{i=4,5} (R_{i})^{2} \sigma_{i} \delta^{i}.
\end{align}
In the fermionic part of the action (\ref{action of IKKT model}),

\begin{align}
  \label{Dirac ooerator in IKKT model}
  \slashed{D}_{6}=\sum_{i=4}^{9} \Gamma^{i} \left[ \hat{X}_{i}, \cdot \right]
\end{align}
is identified as a Dirac operator on the extra dimensions, e.g., \cite{Nishimura:2013moa}. As mentioned above, we focus on the first two extra dimensions that are the NC torus (i.e., $\sum_{i=4}^{9} \rightarrow \sum_{i=4,5}$). We can verify that eqs. (\ref{Dirac operator consistent with IKKT model}) and (\ref{Dirac ooerator in IKKT model}) are equivalent. Namely,

\begin{align*}
  \slashed{D}_{\text{phys}} \hat{\psi}&=2\pi i \sum_{i=4,5} (R_{i})^{2} \sigma_{i} \delta^{i} \hat{\psi}\\
  &=i\sigma_{i} \theta^{ij} \left(-i (\theta^{-1})_{jk} \left[\hat{X}^{k}, \hat{\psi} \right]\right)\\
  &=\sigma_{i} \left[\hat{X}^{i}, \hat{\psi} \right]\\
  &=\slashed{D}_{6} \hat{\psi}.
\end{align*}
Therefore, the operator (\ref{Dirac operator consistent with IKKT model}) is a suitable Dirac operator on the NC torus based on the IKKT model.

The partial derivatives are defined by eq. (\ref{partial derivative NC torus}), and the gauge field is introduced by eq. (\ref{fluctuation}). The covariant derivatives are naturally defined by

\begin{align}
  \label{covariant derivative}
  \partial_{M} \rightarrow D_{M}:=-i (\theta^{-1})_{M N} \left[\hat{X}^{N}_{\text{bg}}+\theta^{NL}\hat{A}_{L}, \cdot \right]=\partial_{M}-i \left[\hat{A}_{M}, \cdot \right].
\end{align}
However, we cannot introduce any fermions $\hat{\psi}$ in a fundamental representation, i.e., $\hat{\psi} \rightarrow \hat{\Omega} \hat{\psi}$, where $\hat{\Omega}$ is a unitary operator as a gauge transformation. This comes from eq. (\ref{unitary transformation}). On the other hand, we can introduce fermions in a fundamental representation if we define the covariant derivatives as 

\begin{align}
  \label{covariant derivative fundamental rep}
  D_{M}:= \partial_{M}-i\hat{A}_{M}.
\end{align}
However, the original action (\ref{action of IKKT model}) does not have the field strength which is defined by the covariant derivatives (\ref{covariant derivative fundamental rep}). Therefore, if we want to consider fermions in a fundamental representation, we should realize as a part of the action (\ref{action of IKKT model}). Let us consider, for example, $U(2)$ gauge theory with an adjoint matter based on the action (\ref{action of IKKT model}), i.e.,

\begin{align*}
  \hat{X}^{M}_{U(2)}:= \hat{X}^{M}_{U(1)} \times 1_{2 \times 2}, \quad 
  \hat{A}_{M}=
  \begin{pmatrix}
    \hat{A}^{11}_{M} & \hat{A}^{12}_{M} \\
    \hat{A}^{21}_{M} & \hat{A}^{22}_{M}
  \end{pmatrix}, 
  \quad 
  \hat{\psi}=
  \begin{pmatrix}
    \hat{\psi}^{11} & \hat{\psi}^{12} \\
    \hat{\psi}^{21} & \hat{\psi}^{22}
  \end{pmatrix},
\end{align*}
where $\hat{X}^{M}_{(U(1)}$ satisfies $\left[\hat{X}^{M}_{(U(1)}, \hat{X}^{N}_{(U(1)} \right]=i\theta^{M N}$. We assume that the gauge group breaks such that $U(2) \rightarrow U(1) \times U(1) \rightarrow U(1)$: top left component of $\hat{A}_{i}$. If we focus on only $\hat{\psi}^{12}$ and ignore other components of $\hat{A}_{i}$, then we can derive the action with the fermion in $U(1)$ fundamental representation.

\section{Zero-mode analysis on the magnetized NC torus}
\label{5}
\subsection{Twisted bundle on the NC torus}
To consider the twisted bundle on the NC torus, we introduce a background gauge field

\begin{align}
  \label{gauge background on NC torus}
  \hat{A}_{4}(\hat{X}^{4},\hat{X}^{5})=0, \quad \hat{A}_{5}(\hat{X}^{4},\hat{X}^{5})=\mathcal{F} \hat{X}^{4},
\end{align}
and the field strength is $\hat{F}_{45}=\mathcal{F}$.

The gauge transformation is obtained by

\begin{align*}
  \hat{A}_{M} \rightarrow \hat{A}^{'}_{M}=\hat{\Omega} \hat{A}_{M} \hat{\Omega}^{-1}+i\hat{\Omega} \partial_{M} \hat{\Omega}^{-1},
\end{align*}
where $\hat{\Omega}$ is a unitary operator.

The background gauge field is varied by the torus translations (\ref{compactification conditions}), i.e.,

\begin{align}
  \label{gauge transformation under NC torus translations}
  \hat{A}_{4}(\hat{X}^{4}+2 \pi R_{4}, \hat{X}^{5})&=\hat{A}_{4}(\hat{X}^{4}, \hat{X}^{5}+2 \pi R_{5})=0, \notag \\
  \hat{A}_{5}(\hat{X}^{4}+2 \pi R_{4}, \hat{X}^{5})&=\hat{A}_{5}(\hat{X}^{4}, \hat{X}^{5})+2\pi R_{4} \mathcal{F}, \notag \\
  \hat{A}_{5}(\hat{X}^{4}, \hat{X}^{5}+2\pi R_{5})&=\hat{A}_{5}(\hat{X}^{4}, \hat{X}^{5}).
\end{align}
We can realize eq. (\ref{gauge transformation under NC torus translations}) as the gauge transformations, i.e.,

\begin{align}
  \label{gauge transition function for NC torus translations}
  \hat{\Omega}_{4}(\hat{X}^{4},\hat{X}^{5}) \propto \exp \left[ \frac{2\pi i R_{4}}{1+\theta^{45} \mathcal{F}} \cdot \mathcal{F} \hat{X}^{5} \right], \quad \hat{\Omega}_{5}(\hat{X}^{4},\hat{X}^{5}) \propto \hat{1},
\end{align}
where $\propto$ represents an action on $\mathbb{C}^{m}$ part. This part does not affect eq. (\ref{gauge transformation under NC torus translations}) since the $\mathbb{C}^{m}$ part of $\hat{X}^{4}$ and $\hat{X}^{5}$ is the identity matrix. Therefore, in the following, we assume this part is the identity matrix and omit.

Next, we must consider the consistency condition corresponding to eq. (\ref{consistency condition on torus}). Namely,

\begin{align}
  \label{consistency condition on NC torus}
  \hat{\Omega}_{5}(\hat{X}^{4}+2 \pi R_{4}, \hat{X}^{5}) \hat{\Omega}_{4}(\hat{X}^{4},\hat{X}^{5})=\hat{\Omega}_{4}(\hat{X}^{4},\hat{X}^{5}+2\pi R_{5}) \hat{\Omega}_{5}(\hat{X}^{4},\hat{X}^{5}).
\end{align}
Eq. (\ref{consistency condition on NC torus}) implies that the magnetic flux $\mathcal{F}$ is quantized such that 

\begin{align}
  \label{quantization condition on NC torus}
  \frac{\mathcal{F}}{1+\theta^{45} \mathcal{F}} \cdot \frac{A}{2\pi}=\mathcal{N} \in \mathbb{Z}.
\end{align}

\subsection{Zero-modes of the Dirac operator $\slashed{D}_{\text{phys}}$}
In this subsection, we consider the zero-mode equations of the Dirac operator $\slashed{D}_{\text{phys}}$ with the background gauge field (\ref{gauge background on NC torus}). In the following, we show the zero-mode solutions in (i) fundamental representation (ii) bifundamental representation.

\begin{itemize}
  \item{Case (i)}
\end{itemize}
In this case, the zero-mode equation is 

\begin{align*}
  \slashed{D}_{\text{phys}} \hat{\psi} =\theta^{45} 
  \begin{pmatrix}
    0 & -(\partial_{4}-i\partial_{5}-\mathcal{F}\hat{X}^{4}) \\
    \partial_{4}+i\partial_{5}+\mathcal{F}\hat{X}^{4} & 0
  \end{pmatrix}
  \begin{pmatrix}
    \hat{\psi}^{+} \\
    \hat{\psi}^{-}
  \end{pmatrix}
  =0,
\end{align*}
where $\hat{\psi}^{s}$ $(s=\pm 1)$ is in the fundamental representation. Simply,

\begin{align}
  \label{NC zero-mode equation fundamental rep}
  (\partial_{4}+is\partial_{5}+s\mathcal{F}\hat{X}^{4}) \hat{\psi}^{s}=0 \quad (s=\pm 1)
\end{align}
with the periodic boundary conditions

\begin{align*}
  &\hat{\psi}^{s}(\hat{X}^{4}+2\pi R_{4},\hat{X}^{5})=\exp \left[\frac{2\pi i R_{4}}{1+\theta^{45}\mathcal{F}} \cdot \mathcal{F} \hat{X}^{5} \right] \hat{\psi}^{s}(\hat{X}^{4},\hat{X}^{5})\\
  & \hat{\psi}^{s}(\hat{X}^{4},\hat{X}^{5}+2\pi R_{5})= \hat{\psi}^{s}(\hat{X}^{4},\hat{X}^{5})
\end{align*}
Although eq. (\ref{NC zero-mode equation fundamental rep}) is written by the operators, this equation has the same form with eq. (\ref{zero-mode equation each component on torus}). Therefore, we expect that the zero-mode equation (\ref{NC zero-mode equation fundamental rep}) can be constructed from that of eq. (\ref{zero-mode equation each component on torus}). Here, we consider the Fourier transformations. The Fourier transformation of the whole zero-mode solution of eq. (\ref{zero-mode equation each component on torus}) is obtained by

\begin{align}
  \label{fourier transformation of zero-mode solution on torus}
  \psi^{s}(x^{4},x^{5})=\int \frac{dk}{2 \pi} \sum_{n=|N|p+q \in \mathbb{Z}} C \exp \left[-\frac{k^{2}}{2|F|}-\frac{ikn}{R_{5}F} \right] \exp \left[ikx^{4}\right] \exp \left[i \frac{n}{R_{5}}x^{5} \right],
\end{align}
where $C$ is the normalization constant and the magnetic flux $F$ satisfies the quantization condition (\ref{quantization of magnetic flux}). We expect that the whole solution of eq. (\ref{NC zero-mode equation fundamental rep}) is obtained by

\begin{align}
  \label{incorrect zero-mode solution NC torus}
  \hat{\psi}^{s} (\hat{X}^{4},\hat{X}^{5}) =\int \frac{dk}{2\pi} \sum_{n=|N|p+q  \in \mathbb{Z}} C \exp \left[-\frac{k^{2}}{2|F|}-\frac{ikn}{R_{5}F} \right] \exp \left[ik\hat{X}^{4}\right] \exp \left[i \frac{n}{R_{5}} \hat{X}^{5} \right].
\end{align}
However, eq. (\ref{incorrect zero-mode solution NC torus}) does not satisfy eq. (\ref{NC zero-mode equation fundamental rep}) because the quantization conditions for the magnetic flux are different.

Here, we focus on that $p$ and $q$ do not appear alone in eq. (\ref{fourier transformation of zero-mode solution on torus}). This fact allows us to vary the decomposition of the label from $n=|N|p+q$ to $n=|\mathcal{N}|p+q$. In addition, we should replace the magnetic flux $F$ with $\mathcal{F}$ satisfying (\ref{quantization condition on NC torus}). Then, we can obtain the zero-mode solutions\footnote{We can obtain the same result by using the Baker-Campbell-Hausdorff formula.}, i.e.,

\begin{align*}
  \hat{\psi}^{s}(\hat{X}^{4},\hat{X}^{5}) = \int \frac{dk}{2\pi} \sum_{n=|\mathcal{N}|p+q  \in \mathbb{Z}} C \exp \left[-\frac{k^{2}}{2|\mathcal{F}|}-\frac{ikn}{R_{5}\mathcal{F}}\right] \exp \left[ik\hat{X}^{4}\right] \exp \left[ i \frac{n}{R_{5}} \hat{X}^{5}\right],
\end{align*}
or

\begin{align}
  \label{each fundamental zero-mode solutions on NC torus}
  \hat{\psi}^{s}_{q}(\hat{X}^{4},\hat{X}^{5}) = \int \frac{dk}{2\pi} \sum_{p \in \mathbb{Z}} C \exp \left[-\frac{k^{2}}{2|\mathcal{F}|}-\frac{ik}{R_{5}\mathcal{F}}(|\mathcal{N}|p+q)\right] \exp \left[ik\hat{X}^{4}\right] \exp \left[ i \frac{|\mathcal{N}|p+q}{R_{5}} \hat{X}^{4}\right],
\end{align}
where $q=0 \sim |\mathcal{N}|-1$. We can confirm that the above expansions satisfy eq. (\ref{NC zero-mode equation fundamental rep}) if $\mathcal{F}s>0$. The normalization constant $C$ will be computed in subsection \ref{5.4}.

However, this form is difficult to use when we compute the normalization constant and Yukawa couplings. Fortunately, we can rewrite by using the Jacobi theta-function

\begin{align}
  \label{fundamental zero-mode solution theta function form on NC torus}
  \hat{\psi}^{s}_{q} (\hat{X}^{4},\hat{X}^{5}) =C \sqrt{\frac{|\mathcal{F}|}{2\pi}} \exp \left[-\frac{|\mathcal{F}|}{2} (\hat{X}^{4})^{2} \right]
  \vartheta
  \begin{bmatrix}
    q/\mathcal{N} \\
    0
  \end{bmatrix}
  \left(\frac{s|\mathcal{N}|}{2\pi i R_{5}}(\hat{X}^{4}+is\hat{X}^{5}), i \frac{R_{4}}{R_{5}} s \mathcal{N}\right).
\end{align}
If we require $s \mathcal{N}>0$, eq. (\ref{fundamental zero-mode solution theta function form on NC torus}) can be interpreted as the operator form of eq. (\ref{normalized zero-mode solution torus}) up to the normalization constant. Therefore, in the following, we restrict ourselves to the case $s \mathcal{N}>0$.

\begin{itemize}
  \item{Case(ii)}
\end{itemize}
For simplicity, we consider the magnetic flux breaks the gauge group $U(2) \rightarrow U(1) \times U(1)$, i.e.,

\begin{align*}
  \hat{A}_{4}(\hat{X}^{4},\hat{X}^{5})=0, \quad \hat{A}_{5}(\hat{X}^{4},\hat{X}^{5})=
  \begin{pmatrix}
    \mathcal{F}_{1} \hat{X}^{4} & 0\\
    0 & \mathcal{F}_{2} \hat{X}^{5}
  \end{pmatrix},
\end{align*}
where $\mathcal{F}_{i}$ $(i=1,2)$ satisfies (\ref{quantization condition on NC torus}).

In this case, we consider a fermion in $U(2)$ adjoint representation,

\begin{align*}
  \hat{\psi}=
  \begin{pmatrix}
    \hat{\psi}^{+}\\
    \hat{\psi}^{-}
  \end{pmatrix},
  \quad 
  \hat{\psi}^{s}=
  \begin{pmatrix}
    \hat{\psi}^{s}_{11} & \hat{\psi}^{s}_{12}\\
    \hat{\psi}^{s}_{21} & \hat{\psi}^{s}_{22}
  \end{pmatrix}.
\end{align*}
The zero-mode equation for each chirality is written as 

\begin{align*}
  (\partial_{4}+is\partial_{5}) 
  \begin{pmatrix}
    \hat{\psi}^{s}_{11} & \hat{\psi}^{s}_{12}\\
    \hat{\psi}^{s}_{21} & \hat{\psi}^{s}_{22}
  \end{pmatrix}
  +s
  \begin{pmatrix}
    \mathcal{F}_{1} [\hat{X}^{4},\hat{\psi}^{s}_{11}] & \mathcal{F}_{1} \hat{X}^{4} \hat{\psi}^{s}_{12}- \mathcal{F}_{2} \hat{\psi}^{s}_{12} \hat{X}^{4} \\
    \mathcal{F}_{2} \hat{X}^{4} \hat{\psi}^{s}_{21}- \mathcal{F}_{1} \hat{\psi}^{s}_{21} \hat{X}^{4} & \mathcal{F}_{2} [\hat{X}^{4},\hat{\psi}^{s}_{22}] 
  \end{pmatrix}
  =0 \quad (s=\pm 1).
\end{align*}

First, we consider the top right component $\hat{\psi}^{s}_{12}$. Precisely, the zero-mode equation of $\hat{\psi}^{s}_{12}$ is

\begin{align}
  \label{NC zero-mode equation bifundamental}
  (\partial_{4}+is \partial_{5}) \hat{\psi}^{s}_{12} +s(\mathcal{F}_{1} \hat{X}^{4} \hat{\psi}^{s}_{12} - \mathcal{F}_{2} \hat{\psi}^{s}_{12} \hat{X}^{4})=0
\end{align}
with the periodic boundary conditions

\begin{align*}
  &\hat{\psi}^{s}_{12}(\hat{X}^{4}+2\pi R_{4},\hat{X}^{5})=\exp \left[ \frac{2\pi i R_{4}}{1+\theta^{45} \mathcal{F}_{1}} \cdot \mathcal{F}_{1} \hat{X}^{5} \right]\hat{\psi}^{s}_{12}(\hat{X}^{4},\hat{X}^{5}) \exp \left[ -\frac{2\pi i R_{4}}{1+\theta^{45} \mathcal{F}_{2}} \cdot \mathcal{F}_{2} \hat{X}^{5} \right] \\
  &\hat{\psi}^{s}_{12}(\hat{X}^{4}, \hat{X}^{5}+2\pi R_{5} )=\hat{\psi}^{s}_{12}(\hat{X}^{4},\hat{X}^{5})
\end{align*}
In the case (i), $\exp \left[-\frac{|\mathcal{F}|}{2} (\hat{X}^{4})^{2} \right]$ of eq. (\ref{fundamental zero-mode solution theta function form on NC torus}) corresponds to the term: $s \mathcal{F} \hat{X}^{4}$ in eq. (\ref{NC zero-mode equation fundamental rep}). From this observation, we can obtain the solutions of eq. (\ref{NC zero-mode equation bifundamental}) as

\begin{align*}
  \hat{\psi}^{s}_{12}(\hat{X}^{4},\hat{X}^{5}) = \int \frac{dk}{2\pi } \sum_{n=|\mathcal{N}_{12}|p+q \in \mathbb{Z} } C_{12} \exp \left[ -\frac{k^{2}}{2|\mathcal{F}_{12}|}-\frac{ikn}{R_{5}\mathcal{F}_{12}}(1+\theta^{12}\mathcal{F}_{2})  \right] \exp \left[ik \hat{X}^{4} \right] \exp \left[i \frac{n}{R_{5}} \hat{X}^{5} \right],
\end{align*}
or

\begin{align}
  \label{bifundamental zero-mode solution theta function}
  \hat{\psi}^{s}_{12,q} (\hat{X}^{1},\hat{X}^{2}) = C_{12} \sqrt{ \frac{|\mathcal{F}_{12}|}{2\pi} } &\exp \left[ -\frac{s_{12} \mathcal{F}_{1} }{2} (\hat{X}^{1})^{2} \right] \notag \\
  & \times \vartheta
  \begin{bmatrix}
    q/|\mathcal{N}_{12}| \\
    0
  \end{bmatrix}
  \left(\frac{s|\mathcal{N}_{12}|}{2\pi i R_{5}} (\hat{X}^{1}+is\hat{X}^{2}), i \frac{R_{4}}{R_{5}} |\mathcal{N}_{12}| \right)
  \exp \left[ \frac{s_{12}\mathcal{F}_{2}}{2} (\hat{X}^{1})^{2} \right]
\end{align}
where $\mathcal{F}_{12}=\mathcal{F}_{1}-\mathcal{F}_{2}$, $s_{12}=\text{sign}(\mathcal{F}_{12})$, $\mathcal{N}_{12}= \mathcal{N}_{1}-\mathcal{N}_{2}$, and we restrict ourselves to $s_{12} \mathcal{N}_{12}$. We can confirm that eq. (\ref{bifundamental zero-mode solution theta function}) satisfies eq. (\ref{NC zero-mode equation bifundamental}) and the periodic boundary conditions if $s_{12} \mathcal{F}_{12}>0$.

Next, we consider the top left component $\hat{\psi}^{s}_{11}$. The zero-mode equation of $\hat{\psi}^{s}_{11}$ is written as

\begin{align*}
  (\partial_{4}+is\partial_{5}) \hat{\psi}^{s}_{11}+s\mathcal{F}_{1} [\hat{X}^{4},\hat{\psi}^{s}_{11}]=0
\end{align*}
with the periodic boundary conditions

\begin{align}
  \label{pbc of diagonal component NC torus}
  &\hat{\psi}^{s}_{11}(\hat{X}^{4}+2\pi R_{4},\hat{X}^{5})= \exp \left[ \frac{2\pi i R_{4}}{1+\theta^{45}\mathcal{F}_{1}} \cdot \mathcal{F}_{1} \hat{X}^{5} \right] \hat{\psi}^{s}_{11}(\hat{X}^{4},\hat{X}^{5}) \exp \left[- \frac{2\pi i R_{4}}{1+\theta^{45}\mathcal{F}_{1}} \cdot \mathcal{F}_{1} \hat{X}^{5} \right], \notag \\
  &\hat{\psi}^{s}_{11}(\hat{X}^{4},\hat{X}^{5}+2\pi R_{5})= \hat{\psi}^{s}_{11}(\hat{X}^{4},\hat{X}^{5}).
\end{align}
In this case, the right-hand side of the first condition of (\ref{pbc of diagonal component NC torus}) corresponds to the shift of $\hat{X}^{4}$ such that $\hat{X}^{4} \rightarrow \hat{X}^{4} + \frac{\mathcal{N}_{1}\theta^{45}}{R_{5}}$. This shift implies $\frac{\mathcal{F}_{1}\theta^{45}}{1+\theta^{45}\mathcal{F}_{1}}=1$, and this is a contradiction. Therefore, $\hat{\psi}^{s}_{11}$ must be the scalar operator, i.e., $\hat{\psi}^{s}_{11}=\text{const.} \times \mathbf{1}$.

Obviously, the above results can be generalized to the magnetic fluxes break the gauge group $U(N) \rightarrow \prod_{a=1}^{n} U(N_{a})$, where $\sum_{a=1}^{n} N_{a}=N$.

\subsection{Eigenvalues of the Laplacian}
In the IKKT model, $\Box:= \sum_{i} [\hat{X}^{i},[\hat{X}^{i}, \cdot]]$ (in this paper, $\Box:= \sum_{i=4,5} [\hat{X}^{i},[\hat{X}^{i}, \cdot]]$) is identified as the Laplacian (e.g., \cite{Chatzistavrakidis:2011gs,Steinacker:2014fja} or eq. (\ref{app Abelian magnetic flux})). The eigenvalue problem of the Laplacian relates to that of the square of the Dirac operator. Let us rewrite the Dirac operator as

\begin{align*}
  \slashed{D}_{\text{phys}}=
  \begin{pmatrix}
    0 & -D \\
    D' & 0
  \end{pmatrix},
\end{align*}
then

\begin{align}
  \label{square of Dirac operator}
  \slashed{D}^{2}_{\text{phys}}=
  \begin{pmatrix}
    -DD' & 0\\
    0 -D'D
  \end{pmatrix}
  =\Box+
  \begin{pmatrix}
    -\frac{(\theta^{45})^{2}}{2}[D,D'] & 0\\
    0 & -\frac{(\theta^{45})^{2}}{2}[D',D]
  \end{pmatrix}.
\end{align}
The action of (\ref{square of Dirac operator}) on $(1,2)$ component of the fermion is written as

\begin{align*}
  \slashed{D}^{2}_{\text{phys}} \hat{\psi}^{s}_{12}=\Box \hat{\psi}^{s}_{12}-s(\theta^{45})^{2} \mathcal{F}_{12} \hat{\psi}^{s}_{12}.
\end{align*}
From the above, $D'$ $(D)$ has the zero-mode solutions if $\mathcal{F}_{12}>0$ $(\mathcal{F}_{12}<0)$. Therefore, we can see that eq. (\ref{bifundamental zero-mode solution theta function}) is not only the zero-modes of the Dirac operator but also the lightest mode of the Laplacian.

In addition, we can construct the eigenmodes of the Laplacian corresponding $(1,2)$ component. We focus on the commutation relation of $D$ and $D'$,

\begin{align*}
  [D,D'] \hat{\psi}^{s}_{12}=2(\theta^{45})^{2} \mathcal{F}_{12} \hat{\psi}^{s}_{12}.
\end{align*}
If we select $\mathcal{F}_{12}>0$, $D$ and $-D'$ can play roles of the creation operator and the annihilation operator, respectively. By considering the analogy of the harmonic oscillator, i.e.,

\begin{align*}
  N:=-DD', \quad \Box=N+(\theta^{45})^{2} \mathcal{F}_{12},
\end{align*}
then the eigenmodes of the Laplacian are obtained by 

\begin{align*}
  &\Box \hat{\psi}^{+,n}_{12}=\lambda_{n} \hat{\psi}^{+,n}_{12},\\
  &\hat{\psi}^{+,n}_{12}:= D^{n} \hat{\psi}^{+}_{12}, \quad \lambda_{n}=(\theta^{45})^{2} \mathcal{F}_{12} (2n+1).
\end{align*}
If we select $\mathcal{F}_{12}<0$, it is sufficient to reverse the roles of $D$ and $D'$.

In the commutative case\footnote{We use {\it commutative} in the sense of usual torus $T^{2}$, not the case: $[X^{M},X^{N}]=0$ $(\text{for all } M,N=0 \sim 9)$.}, the spectrum of the Laplacian is obtained by $\lambda_{n}=F_{12}(2n+1)$, where $F_{i}$ $(i=1,2)$ is the magnetic flux satisfying (\ref{quantization of magnetic flux}) (c.f., \cite{Cremades:2004wa}) and $F_{12}=F_{1}-F_{2}$. It seems that the spectrum vanishes in the limit $\theta^{45} \rightarrow 0$. However, the bosonic part of the effective action is the fourth order for the NC parameter if we ignore the order included in the definition of the partial derivatives. Therefore, $(\theta^{45})^{2}$ of the spectrum is necessary, but the limit $\theta^{45} \rightarrow 0$ is non-trivial.

\subsection{Normalizations and Yukawa couplings}
\label{5.4}

\begin{itemize}
  \item {Normalizations}
\end{itemize}

Let us start from the normalization constant of eq. (\ref{bifundamental zero-mode solution theta function}) since $\mathcal{F}_{2}=0$ corresponds to the case (i).

\begin{align*}
  1=\text{Tr} \left( \hat{\psi}^{s,\dagger}_{12,q}(\hat{X}^{4},\hat{X}^{5}) \hat{\psi}^{s}_{12, q'}(\hat{X}^{4},\hat{X}^{5})  \right)&:= \int_{0}^{2\pi R_{5}} dX^{5} \Bra{X^{5}}  \hat{\psi}^{s,\dagger}_{12,q}(\hat{X}^{4},\hat{X}^{5}) \hat{\psi}^{s}_{12, q'}(\hat{X}^{4},\hat{X}^{5}) \Ket{X^{5}}\\
  &=\frac{1}{A} \int_{0}^{2\pi R_{4}} dX^{4} \int_{0}^{2\pi R_{5}} dX^{5} \hat{\psi}^{s,\dagger}_{12,q}(X^{4}, X^{5}) \hat{\psi}^{s}_{12, q'}(X^{4},X^{5})\\
  &=\delta_{qq'} |C_{12}|^{2} \frac{R_{5}|\mathcal{F}_{12}|}{A} \cdot I_{12}(\theta^{12}),
\end{align*}
where

\begin{align*}
  I_{12}(\theta^{12}):= \sum_{p \in \mathbb{Z}} \int_{\frac{p|\mathcal{N}_{12}|p+q}{R_{5}\mathcal{F}_{12}}(1+\theta^{45}\mathcal{F}_{2})}^{ \frac{p|\mathcal{N}_{12}|p+q}{R_{5}\mathcal{F}_{12}}(1+\theta^{45}\mathcal{F}_{2})  -2\pi R_{4}} \exp \left[-|\mathcal{F}_{12}|x^{2}\right] dx
\end{align*}
is a function satisfies $I_{12}(0)=\sqrt{\pi/|\mathcal{F}_{12}|}$ (the Gaussian integral). Therefore, the normalization constant $C_{12}$ is 

\begin{align*}
  C_{12}= \left(\frac{R_{5}|\mathcal{F}_{12}|}{A} \cdot I_{12} (\theta^{12})\right)^{-1/2}.
\end{align*}

We should note the definition of the trace on the infinite dimensional space. For example,

\begin{align}
  \label{trace x1 0-2 pi R1}
  \int_{0}^{2\pi R_{4} } dX^{4} \Bra{X^{4}} \hat{f}(\hat{X}^{4},\hat{X}^{5}) \Ket{X^{4}} &= \int_{0}^{2\pi R_{4} } dX^{4} \int_{0}^{2\pi R_{5}} dX^{5} \Bra{X^{4}} \hat{f}(\hat{X}^{4},\hat{X}^{5}) \Ket{X^{5}} \Braket{X^{5}|X^{4}} \notag \\
  &=\int_{0}^{2\pi R_{4} } dX^{4} \int_{0}^{2\pi R_{5}} dX^{5} \Braket{X^{5}|X^{4}} \Bra{X^{4}} \hat{f}(\hat{X}^{4},\hat{X}^{5}) \Ket{X^{5}} \notag \\
  &=\int_{0}^{2\pi R_{5}} dX^{5} \Bra{X^{5}} \hat{f}(\hat{X}^{4},\hat{X}^{5}) \Ket{X^{5}} \notag \\
  &=\text{Tr} \left( \hat{f}(\hat{X}^{4},\hat{X}^{5}) \right).
\end{align}
In addition, we should confirm that the trace is well-defined with respect to the compactification conditions (\ref{compactification conditions}). In this paper, target operators $\hat{f}$ are periodic with respect to the $X^{5}$-direction. This means

\begin{align*}
  \int_{2\pi R_{5}}^{4\pi R_{5}} dX^{5} \Bra{X^{5}} \hat{f}(\hat{X}^{4},\hat{X}^{5}) \Ket{X^{5}} &= \int_{0}^{2\pi R_{5}} dX^{5} \Bra{X^{5}+2\pi R_{5}} \hat{f}(\hat{X}^{4},\hat{X}^{5}) \Ket{X^{5}+2\pi R_{5}}\\
  &= \int_{0}^{2\pi R_{5}} dX^{5} \Bra{X^{5}} \hat{U}_{5} \hat{f}(\hat{X}^{4},\hat{X}^{5}) \hat{U}^{-1}_{5} \Ket{X^{5}}\\
  &=\int_{0}^{2\pi R_{5}} dX^{5} \Bra{X^{5}} \hat{f} (\hat{X}^{4},\hat{X}^{5}+2\pi R_{5}) \Ket{X^{5}}\\
  &=\int_{0}^{2\pi R_{5}} dX^{5} \Bra{X^{5}} \hat{f} (\hat{X}^{4},\hat{X}^{5}) \Ket{X^{5}}\\
  &=\text{Tr} \left( \hat{f}(\hat{X}^{4},\hat{X}^{5}) \right).
\end{align*}
On the other hand, the target operators $\hat{f}$ are quasiperiodic with respect to the $X^{4}$-direction. Since the unitary operator for the quasiperiodicity $\hat{\Omega}_{4}$ is written by $\hat{X}^{5}$ ($\hat{\Omega}_{4}$ is also periodic with respect to the $X^{5}$-direction), then

\begin{align*}
  \int_{2\pi R_{4}}^{4\pi R_{4}} dX^{4} \Bra{X^{4}} \hat{f}(\hat{X}^{4},\hat{X}^{5}) \Ket{X^{4}} &= \int_{0}^{2\pi R_{4}} dX^{4} \Bra{X^{4}+2\pi R_{4}} \hat{f}(\hat{X}^{4},\hat{X}^{5}) \Ket{X^{4}+2\pi R_{4}}\\
  &=\int_{0}^{2\pi R_{4}} dX^{4} \Bra{X^{4}} \hat{U}_{4} \hat{f}(\hat{X}^{4},\hat{X}^{3})  \hat{U}^{-1}_{4} \Ket{X^{4}}\\
  &=\int_{0}^{2\pi R_{4}} dX^{4} \Bra{X^{4}} \hat{\Omega}_{4} \hat{f}(\hat{X}^{4},\hat{X}^{5})  \hat{\Omega}^{-1}_{4} \Ket{X^{4}}\\
  &=\int_{0}^{2\pi R_{5}} dX^{5} \Bra{X^{5}} \hat{\Omega}_{4} \hat{f}(\hat{X}^{4},\hat{X}^{5})  \hat{\Omega}^{-1}_{4} \Ket{X^{5}}\\
  &=\text{Tr} \left( \hat{f}(\hat{X}^{4},\hat{X}^{5}) \right),
\end{align*}
where we used the equivalence (\ref{trace x1 0-2 pi R1}) in the fourth line.

For general gauge transformations $\hat{U}$, we can show the equivalence between before and after the gauge transformations if we assume the existence of the completeness relation of $\hat{U}$. Therefore, the cyclic property of the trace is held, at least, for the gauge transformations. Therefore, the gauge symmetry of the action (\ref{action of IKKT model}) is still held. Similarly, we can verify that all traces defined by the appropriate completeness relation are equivalent.

\begin{itemize}
  \item{Yukawa couplings}
\end{itemize}
In the following, for simplicity, we consider the magnetic fluxes break the gauge group $U(N) \rightarrow \prod_{a=1}^{3} U(N_{a})$, where $\sum_{a=1}^{3}N_{a}=N$ to compute the Yukawa couplings. The background gauge field is obtained by 

\begin{align*}
  \hat{A}_{4}(\hat{X}^{5},\hat{X}^{5})=0, \quad \hat{A}_{5}(\hat{X}^{4},\hat{X}^{5})=
  \begin{pmatrix}
    \mathcal{F}_{1} \hat{X}^{4} \mathbf{1}_{N_{1}} & 0 & 0 \\
    0 & \mathcal{F}_{2} \hat{X}^{4} \mathbf{1}_{N_{2}} & 0\\
    0 & 0 & \mathcal{F}_{3} \hat{X}^{4} \mathbf{1}_{N_{3}}
  \end{pmatrix},
\end{align*}
where $\mathcal{F}_{i}$ $(i=1,2,3)$ satisfies (\ref{quantization condition on NC torus}) and $\mathbf{1}_{N_{i}}$ is the $N_{i} \times N_{i}$ identity matrix.

We should consider a sign assignment of the magnetic fluxes. In the following, we select $\mathcal{F}_{23},\mathcal{F}_{21},\mathcal{F}_{13}>0$ (This implies $s_{23},s_{21},s_{13}=+1$ and $\mathcal{N}_{23},\mathcal{N}_{21},\mathcal{N}_{13}>0$). This sign assignment is justified by the relation $\mathcal{F}_{12}+\mathcal{F}_{23}+\mathcal{F}_{31}=0$.

The component of the zero-mode fermions corresponding the above sign assignment is obtained by 

\begin{align*}
  \hat{\psi}=
  \begin{pmatrix}
    \hat{\psi}^{+}\\
    \hat{\psi}^{-}
  \end{pmatrix},
  \quad 
  \hat{\psi}^{+}=
  \begin{pmatrix}
    \text{const.} & 0 & \hat{\psi}^{+}_{13,i} \\
    \hat{\psi}^{+}_{21,j} & \text{const.} & \hat{\psi}^{+}_{23,k} \\
    0 & 0 & \text{const.} 
  \end{pmatrix},
  \quad \hat{\psi}^{-}=\hat{\psi}^{+, \dagger},
\end{align*}
where $i,j$ and $k$ denote the degeneracies and $\text{const.}=\mathbf{\hat{1}}$ (corresponds to the wavefunctions of the gauginos in the commutative case). The lightest mode bosons have the same matrix structure with the (anti-) chiral zero-mode fermions.

Let us denote by $\Phi_{ab,I}$ $(a,b=1,2,3)$ the $(a,b)$ block component of the zero-mode fermions or the lightest mode bosons since their function form is the same. In the action (\ref{action of IKKT model}), the Yukawa couplings are described as the product of three matrices, i.e.,

\begin{align}
  \label{definition of Yukawa}
  Y_{IJK}:= \text{Tr} \left(\hat{\Phi}^{\dagger}_{23,K} \cdot \hat{\Phi}_{21,I} \cdot \hat{\Phi}_{13,J}   \right).
\end{align}
First, we focus on the product $\hat{\Phi}_{21,I} \cdot \hat{\Phi}_{13,J}$. If operators commute each other, these operators can be regarded as c-numbers. This allows us to use convenient formulas. From eq. (\ref{bifundamental zero-mode solution theta function}), this product can be written by 

\begin{align*}
  \hat{\Phi}_{21.I} \cdot \hat{\Phi}_{13,J} &=\sqrt{ \frac{2\pi |\mathcal{F}_{21} \mathcal{F}_{13}|}{|\mathcal{F}_{23}|}} \frac{C_{21} C_{13}}{C_{23}} \\ 
  & \hspace{0.5cm} \times \sum_{m \in \mathbb{Z}_{|\mathcal{N}_{21}|+|\mathcal{N}_{13}|}} \hat{\Phi}_{23,I+J+|\mathcal{N}_{21}|m} \times \vartheta
  \begin{pmatrix}
    \frac{|\mathcal{N}_{13}|I-|\mathcal{N}_{21}|J+|\mathcal{N}_{21}||\mathcal{N}_{13}|m}{|\mathcal{N}_{21}\mathcal{N}_{23}\mathcal{N}_{13}|} \\
    0
  \end{pmatrix}
  \left(0, i \frac{R_{4}}{R_{5}} |\mathcal{N}_{21}\mathcal{N}_{23}\mathcal{N}_{13}| \right),
\end{align*}
where we used $s_{21},s_{13}=+1$ and the product formula of the Jacobi theta-function \cite{Mumford:1983}

\begin{align*}
  \vartheta 
  \begin{bmatrix}
    \frac{r_{1}}{ N_{1} } \\ 0
  \end{bmatrix}
  (z_{1},\tau N_{1})
  \cdot
  \vartheta 
  \begin{bmatrix}
    \frac{r_{2}}{ N_{2} } \\ 0
  \end{bmatrix}
  (z_{2},\tau N_{2})
  &=\sum_{m \in \mathbb{Z}_{N_{1}+N_{2}}}  \vartheta 
  \begin{bmatrix}
    \frac{r_{1}+r_{2}+N_{1}m }{ N_{1}+N_{2} }\\ 0
  \end{bmatrix}
  (z_{1}+z_{2},\tau (N_{1}+N_{2}) ) \\
  &\hspace{0.5cm} \times \vartheta
  \begin{bmatrix}
    \frac{N_{2}r_{1}-N_{1}r_{2}+N_{1}N_{2}m }{N_{1}N_{2} (N_{1}+N_{2}) }\\ 0
  \end{bmatrix}
  (z_{1}N_{2}-z_{2}N_{1},\tau N_{1}N_{2}(N_{1}+N_{2}) ).
\end{align*}

On the other hand, the orthogonality of $\hat{\Phi}_{ab,I}$ is assured in the above result. Therefore, the Yukawa couplings (\ref{definition of Yukawa}) is obtained by

\begin{align}
  \label{explicit Yukawa}
  Y_{IJK} = \sqrt{ \frac{2\pi A}{R_{5}} \cdot \frac{I_{23}}{I_{21}I_{13}}} \vartheta
  \begin{pmatrix}
    \frac{1}{|\mathcal{N}_{21}|} \left(\frac{K}{|\mathcal{N}_{23}|} - \frac{J}{|\mathcal{N}_{13}|}  \right) \\
    0
  \end{pmatrix}
  \left(0, i \frac{R_{4}}{R_{5}} |\mathcal{N}_{21}\mathcal{N}_{23}\mathcal{N}_{13}|\right),
\end{align}
where we assume $^{\exists}m \in \mathbb{Z}_{|\mathcal{N}_{21}|+|\mathcal{N}_{31}|}$ such that $K=I+J+|\mathcal{N}_{21}|m$.

The Yukawa couplings (\ref{explicit Yukawa}) differ from the commutative case (c.f., \cite{Cremades:2004wa}) by the overall factor only if we fix the generation numbers $\mathcal{N}_{23},\mathcal{N}_{21}$ and $\mathcal{N}_{13}$. On the other hand, we can confirm that the Yukawa couplings (\ref{explicit Yukawa}) go back to the commutative case in the limit $\theta^{45} \rightarrow 0$ since the normalization constant goes back to the commutative case\footnote{The normalization constants are slightly different between the commutative case and our case. This is because the integral of $1$ is normalized to obtain the area of the torus in the commutative case. In our case, the integral of $\hat{1}$ is normalized by considering the analogy of quantum mechanics (\ref{inner product}). This normalization is natural from the viewpoint of the NC torus without magnetic fluxes. Therefore, if we consider the integral of $1$ is normalized to obtain $1$ in the commutative case, the normalization constants are the same between the commutative case and our case.}. However, the limit $\theta^{45} \rightarrow 0$ is non-trivial since the NC parameter $\theta^{45}$ remains in the Yukawa couplings, c.f., $\tilde{\Gamma}^{i}:= \theta^{ij} \Gamma_{j}$ in eq. (\ref{app effective action for fermion}).

\section{Conclusions and discussions}
In this paper, we performed the analysis of the chirality and the generation structures on the magnetized NC torus based on the IKKT model. In subsection \ref{4.2}, we proposed the suitable Dirac operator on the NC torus by considering noncommutative geometry. In section \ref{5}, we analyzed the zero-mode solutions of the Dirac operator we proposed. We showed that zero-mode solutions have the chirality and the generations structures. In addition, we computed the Yukawa couplings of chiral matter fields. Compared with the commutative case, the difference of the Yukawa couplings is the overall factor only. Advantages of our method are (i) we can consider geometric conditions such as periodic boundary conditions (ii) we can write down the analytic form of zero-mode solutions which can easily be compared with the commutative case. This is important to observe NC effects from the IKKT model.

When we consider the microscopic world, we compute the physical quantities through functions on spacetime like wavefunctions. On the other hand, in noncommutative geometry, we consider a function algebra on a certain space which has a NC product. For example, the star-product is a NC product in the context of deformation quantization. In the relationship between analytical mechanics and quantum mechanics, this corresponds to the replacement of the coordinate of the phase space by operators. Therefore, in the sense of noncommutative geometry, we can admit that the chirality and the generation structures of our zero-mode solutions have the physical meanings even though zero-mode solutions are written by the operators.

From a Phenomenological point of view, our results may not be new because the difference of the Yukawa couplings is the overall factor only. However, the important point of this paper is that the IKKT model can describe the string-motivated model including the NC effect. The IKKT model is considered as a non-perturbative formulation of superstring theory. Therefore, we expect that the IKKT model should describe the results of string-motivated models known so far. From this point of view, our results are important.

We are interested in generalizations: orbifolds, complex structure $+$ Wilson line and the origin of the magnetic flux.

The toroidal orbifolds are typical models in string phenomenology. In magnetized toroidal orbifolds, the generation structure differs from the toroidal compactifications without orbifold projections. In Refs. \cite{Konechny:1999zz,Walters:2015bda}, the NC toroidal orbifolds are considered.

On the other hand, in general, the Yukawa couplings in the toroidal compactifications are functions of the complex structure moduli and the Wilson lines. Therefore, the values of the complex structure moduli and the Wilson lines are important to compare with the observed values (c.f., \cite{Abe:2012fj}). Our results correspond to the case whose complex structure $\tau$ is $\tau=iR_{4}/R_{5}$. We can consider the complexification when we introduce the basic derivations. We expect that the Yukawa couplings on the magnetized NC torus are obtained by the general complex structure $\tau$ instead of $\tau=iR_{4}/R_{5}$ and the overall factor including $\tau$.

In our results, the magnetic flux played an important role. However, we introduced the magnetic flux by hand. We expect that the magnetic flux is also generated from the dynamics of the IKKT model. Recently, in Refs. \cite{Asakawa:2017nws,Asakawa:2018gxf}, the authors showed that the magnetic flux may come from the tachyon condensations induced from the dynamics of D-branes and non-BPS D-branes. We expect that our results can be described by the full dynamics of the IKKT model.

\begin{itemize}
  \item {Gauge selection}
\end{itemize}
In this paper, we selected the {\it axial gauge} (\ref{gauge background on NC torus}). Our method, especially the ansatz of the zero-mode solutions which are similar to the Fourier transformation, depends on the gauge selection. Therefore, we should confirm the gauge invariance of our results.

In the commutative case, the background gauge field with the fixed magnetic flux $F$ is obtained by 

\begin{align*}
  A_{4}=-tFx^{5}, \quad A_{5}=(1-t)Fx^{4},
\end{align*}
where $t \in [0,1]$. We can realize the gauge transformation from $^{\forall}t_{1}$ to $^{\forall}t_{2}$ by $U=\exp \left[iF(t_{2}-t_{1})x^{4}x^{5} \right]$.

On the other hand, the background gauge field on the NC torus with the fixed magnetic flux $\mathcal{F}$ is obtained by 

\begin{align*}
  \hat{A}_{4}=-t \mathcal{F} \hat{X}^{5}, \quad \hat{A}_{5}=(1-t) \mathcal{F}\hat{X}^{4},
\end{align*}
where $t \in [0,1]$. We expect that the gauge transformation from $^{\forall}t_{1}$ to $^{\forall}t_{2}$ by the unitary operator, at least, $\hat{U}=\exp \left[i\alpha (\hat{X}^{4}\hat{X}^{5}+\hat{X}^{5}\hat{X}^{4})\right]$, where $\alpha \in \mathbb{R}$. However, we can find a condition such that 

\begin{align}
  \label{gauge selection problem}
  (1-t_{1}\theta^{45}\mathcal{F})(1+(t_{1}-1)\theta^{45}\mathcal{F})=(1-t_{2}\theta^{45}\mathcal{F})(1+(t_{2}-1)\theta^{45}\mathcal{F}) \quad \text{for all } \alpha \in \mathbb{R}.
\end{align}
If we fix the starting point $t=t_{1}$, then $t_{2}$ must be $t_{1}$ or $1-t_{1}$ since eq. (\ref{gauge selection problem}) is a quadratic equation with respect to the $t_{2}$. We need to confirm whether the gauge transformations from $^{\forall}t_{1}$ to $^{\forall}t_{2}$ exist. This situation is the same with a gauge theory on a NC space with the star-product formulation. This is an open question. If we cannot find, different gauge backgrounds may correspond to physically different theories.

\begin{itemize}
  \item {The limit: $\theta^{45} \rightarrow 0$}
\end{itemize}
The zero-mode solutions (\ref{fundamental zero-mode solution theta function form on NC torus}) and (\ref{bifundamental zero-mode solution theta function}) become those of the commutative case if we consider the limit $\theta^{45} \rightarrow 0$ and assume the operators $(\hat{X}^{4},\hat{X}^{5})$ correspond to the coordinate on the torus $(x^{4},x^{5})$. In addition, the Yukawa couplings (\ref{explicit Yukawa}) are the same with the commutative case in this limit. However, from the discussion in Ref. \cite{Steinacker:2007dq} (or Appendix \ref{A}), the NC parameter remains in the effective action, e.g., the effective metric. Therefore, in terms of the effective action, the limit $\theta^{45} \rightarrow 0$ is non-trivial. This is also an open question for us.

\section*{Acknowledgments}
The author would like to thank Hiroyuki Abe for helpful comments.

\def\thesubsection{\Alph{subsection}}
\setcounter{subsection}{0}

\renewcommand{\theequation}{A.\arabic{equation}}
\setcounter{equation}{0}

\section*{Appendix}
\subsection{Effective action of the IKKT model}
\label{A}
We derive the effective action of the IKKT model by considering eq. (\ref{fluctuation}). We refer to \cite{Cremades:2004wa} for the basic techniques. In this Appendix, we consider the whole spacetime again.

The action of the IKKT model is

\begin{align}
  \label{app action of IKKT model}
  S=-\frac{1}{g^{2}} \text{Tr} \left( \frac{1}{4} [\hat{X}_{M},\hat{X}_{N}] [\hat{X}^{M}, \hat{X}^{N}] +\frac{1}{2} \bar{\hat{\psi}} \Gamma^{M} [\hat{X}_{M},\hat{\psi}] \right),
\end{align}
where $M,N=0 \sim 9$. First, we consider the bosonic part of the action (\ref{app action of IKKT model}), i.e.,

\begin{align}
  \label{app bosonic part of the action of IKKT model}
  S_{b}=-\frac{1}{4g^{2}} \text{Tr} \left( [\hat{X}_{M},\hat{X}_{N}] [\hat{X}^{M}, \hat{X}^{N}] \right)
\end{align}
We introduced the gauge field as a fluctuation (\ref{fluctuation}). By substituting, 

\begin{align}
  \label{app bosonic part of the action of IKKT model 2}
  S_{b}=-\frac{1}{4g^{2}} \text{Tr} \left(\eta_{IK} \eta_{JL}  [\hat{X}^{I}_{\text{bg}}+\theta^{I M} \hat{A}_{M}, \hat{X}^{J}_{\text{bg}}+\theta^{J M}\hat{A}_{M}] [\hat{X}^{K}_{\text{bg}}+\theta^{K N} \hat{A}_{N}, \hat{X}^{L}_{\text{bg}}+\theta^{L N} \hat{A}_{N}] \right).
\end{align}
For concreteness, we consider the $U(N)$ gauge group. $(U_{a})^{i}_{j}=\delta_{ai} \delta_{aj}$ and $(e_{ab})_{ij}=\delta_{ai}\delta_{bj}$ can be selected as the basis of the Lie algebra. Accordingly, the trace of the action (\ref{app bosonic part of the action of IKKT model 2}) is defined on the operator space and the gauge group. Then, we can expand the gauge field and the fermions in the adjoint representation as

\begin{align}
  \label{app expansion by lie algebra basis}
  \hat{A}_{M}&=\hat{B}_{M}+\hat{W}_{M}=\hat{B}^{a}_{M}U_{a}+\hat{W}^{ab}_{M}e_{ab}, \notag \\
  \hat{\psi}&=\hat{\chi}+\hat{\Psi}=\hat{\chi}^{a}U_{a}+\hat{\Psi}^{ab}e_{ab}.
\end{align}
Let us define

\begin{align*}
  &\partial_{M}= -i (\theta^{-1})_{M N} [\hat{X}^{N}_{\text{bg}}, \cdot] \quad (\text{Partial derivatives}), \\
  &\hat{F}_{M N}=\partial_{M} \hat{B}_{N}-\partial_{N} \hat{B}_{M}-i[\hat{B}_{M},\hat{B}_{N}] \quad (\text{Field strength of the $U(1)$ gauge group}),\\
  &D_{M}\hat{W}_{N}=\partial_{M} \hat{W}_{N}-i[\hat{B}_{M},\hat{W}_{N}] \quad (\text{Covariant derivatives with respect to the $U(1)$ gauge group}),\\
  &G^{M N}=\theta^{MI} \theta^{NJ} \eta_{I J} \quad (\text{Effective metric}), 
\end{align*}
then eq. (\ref{app bosonic part of the action of IKKT model 2}) is rewritten as

\begin{align}
  \label{app: action of fluctuation}
  S_{b}&=\frac{1}{4g^{2}} \text{Tr} 
    \left(
     \left( \hat{F}_{M N}-(\theta^{-1})_{M N} \right) \left( \hat{F}^{M N}-(\theta^{-1})^{M N} \right) -[\hat{W}_{M},\hat{W}_{N}][\hat{W}^{M},\hat{W}^{N}]
    \right) \notag \\
       &+ \frac{1}{2g^{2}} \text{Tr}
    \left(
      D_{M}\hat{W}_{N} D^{M} \hat{W}^{N} -D_{M} \hat{W}_{N} D^{N} \hat{W}^{M}-i(\hat{F_{M N}} - (\theta^{-1})_{M N})[\hat{W}^{M}, \hat{W}^{N}]  )
    \right),
\end{align}
where the indices are contracted by the effective metric, and we used the cyclic property of the trace.

Next, we introduce the Abelian magnetic flux on the extra-dimensional space $(i=4,5)$, i.e.,

\begin{align}
  \label{app Abelian magnetic flux}
  \hat{B}^{a}_{i} = < \hat{B}^{a}_{i} >+\hat{C}^{a}_{i},\quad \hat{W}^{ab}_{i} =\hat{\Phi}^{ab}_{i},
\end{align}
where $\hat{C}^{a}_{i}$ and $\hat{\Phi}^{ab}_{i}$ are fluctuations around the Abelian magnetic flux background (we set $<\hat{W}^{ab}_{i}>=0$). By substituting the background ($\ref{app Abelian magnetic flux}$), eq. (\ref{app bosonic part of the action of IKKT model 2}) is rewritten as

\begin{align}
  \label{app quadratic terms off diag fluctuation}
  S_{b}=- \frac{1}{2g^{2}} \text{Tr} \left( \hat{\Phi}_{i} D_{\mu} D^{\mu} \hat{\Phi}^{i}+  \hat{\Phi}_{j} \tilde{D}_{i} \tilde{D}^{i} \hat{\Phi}^{j} \right)-\frac{i}{2g^{2}} \left(<\hat{F}^{a}_{ij}> - <\hat{F}^{b}_{ij}> \right) \hat{\Phi}^{i,ab} \hat{\Phi}^{j,ba}+S_{b,\text{other}},
\end{align}
where

\begin{align*}
  \tilde{D}_{i} \hat{\Phi}^{ab}_{j}:=\partial_{i} \hat{\Phi}^{ab}_{j} -i <\hat{B}^{a}_{i}> \hat{\Phi}^{ab}_{j}+i \hat{\Phi}^{ab}_{j} <\hat{B}^{b}_{i}>,
\end{align*}
and we used the cyclic property of the trace. $S_{b,\text{other}}$ contains irrelevant terms for our main discussions.

In eq. (\ref{app quadratic terms off diag fluctuation}), we can see the Laplacian on the extra-dimensional space, i.e.,

\begin{align}
  \label{app Laplacian in IKKT model}
  \Delta_{6d}:= -G^{ij} \tilde{D}_{i} \tilde{D}_{j}=\sum_{i} [\hat{X}^{i},[\hat{X}^{i}, \cdot]].
\end{align}

Similarly, we can obtain the fermionic part of the effective action 

\begin{align}
  \label{app effective action for fermion}
  S_{f}=-\frac{1}{2g^{2}} \text{Tr} \left(i \bar{\hat{\Psi}} \tilde{\Gamma}^{\mu} D_{\mu} \hat{\Psi}+i \bar{\hat{\Psi}} \tilde{\Gamma}^{i} \tilde{D}_{i} \hat{\Psi} + \bar{\hat{\Psi}} \tilde{\Gamma}^{i} [\hat{\Phi}_{i}, \hat{\Psi} ]\right)+S_{f,\text{other}},
\end{align}
where the indices are contracted by the Minkowski metric and $\tilde{\Gamma}^{\mu}:= \theta^{\mu \nu} \Gamma_{\nu}$, and $S_{f,\text{other}}$ contains irrelevant terms for our main discussions.

\bibliographystyle{prsty}

\end{document}